\documentclass[aps,prb,twocolumn,showpacs,superscriptaddress,floatfix]{revtex4-1}
\usepackage{caption}
\usepackage{graphicx} 
\usepackage{subfigure}  
\usepackage{dcolumn}   
\usepackage{bm}  
\usepackage{epstopdf} 
\usepackage{amsmath}      
\usepackage{amssymb}   
\usepackage{bbold}
\usepackage{slashed}
\usepackage{array}
\usepackage{booktabs}
\begin{document}

\title{
Superconductivity From Confinement of Singlets in Metal Oxides
}
\author{J.~M.~Booth}
\email{jamie.booth@rmit.edu.au}
\affiliation{ARC Centre of Excellence in Exciton Science, RMIT University, Melbourne, Australia}

\date{\today}

\pagebreak
\begin{abstract}
The Yang-Mills description of phonons and the consequent structure of electron liquids in strongly anharmonic crystals such as metal oxides is shown to yield an attractive electron-phonon interaction, and thus an instability towards the formation of bound states, which can condense to form a superconductor. This mechanism differs significantly from the pairing mechanism of conventional superconductivity: the ground state from which superconductivity emerges is a many-body state of paired electrons and holes which is not amenable to a quasiparticle description, and whose properties are similar to those seen in the Cuprate high temperature superconductors. Confinement arises because the electron liquid structure acts as a source for Yang-Mills bosons, and not the traditional longitudinal density waves of BCS pairing.
\end{abstract}

\maketitle
The discovery of high temperature superconductivity in the cuprates by Bednorz and M{\"u}ller\cite{Bednorz1986} generated significantly renewed interest in the question of how superconductivity can arise in crystalline materials. In the ensuing three or so decades a huge number of studies have explored both the experimental properties of the cuprates, and their theoretical implications, along the way finding another class of materials which exhibit unconventional superconductivity at elevated temperatures: the pnictides.\cite{Kamihara2006}

However, despite all of this attention, the cuprate phase diagram still contains many mysterious phases, not least of which is the superconducting phase itself,\cite{Mann2011} while the strange-metal\cite{Sachdev2016} and psuedogap\cite{Hashimoto2014} phases are also foci of intense curiosity for condensed matter physics due to the unusual properties they exhibit. The goal of all of this attention is the determination of how high temperature superconductors form, and what sort of ``normal" state they form out of, such that new materials can be engineered which can hopefully produce superconductivity at room-temperature.

There have been many proposals for the underlying mechanism of the emergence of superconductivity in cuprates, beginning with the Resonating Valence Bond\cite{Anderson1987} and spin-wave mediated pairing,\cite{Miyake1986,Scalapino1986} and we cannot begin to summarize them here. For a thorough review of the Mott Insulating perspective see Lee \textit{et al.}\cite{Lee2006}, and for a more general review see Keimer \textit{et al.}\cite{Keimer2015} What is obvious from the literature is that no clear consensus on the pairing mechanism exists apart from i) the recognition that magnetic fluctuations are important, ii) as are strong interactions between the electrons in the normal state, which must be mediated by doping to give a superconducting state, and iii) the fact that whatever the pairing mechanism is, it cannot be the same mechanism which gives rise to pairing in conventional superconductors.\cite{Mann2011}

In two recent studies the application of Yang-Mills theory to condensed matter systems was studied first in the context of electron-phonon interactions,\cite{Booth2018_1} and then a description of anharmonic phonons in terms of pure Yang-Mills theory was presented.\cite{Booth2018_2} In this work, the scattering of Cooper pairs is investigated in the context of a pure Yang-Mills theory coupled to a Fermi surface, which gives an electron phonon mechanism of the type explored in the first paper.\cite{Booth2018_1} It is found that the Yang-Mills scattering cross-section has an attractive potential some spinors, while other scattering events for different spinor combinations are repulsive. The attractive potential occurs for scattering between Cooper pairs.

One simple, but significant assumption is made in this work: the effect of strong correlations in metal oxides is simply to give a tendency towards half-filling, i.e. each electron state is correlated with a hole state, and thus operations on electrons must include the effect on the associated hole. To this end the spinor states which enter the interaction vertex must consist stacked electrons and holes, i.e. Nambu spinors, which are 4-component Weyl spinors. Thus the electrons and holes in a spinor have opposite helicities, and the 4 possible combinations are grouped like so:\cite{Booth2018_1}

\begin{multline}
\begin{pmatrix}\hat{c}^{\dagger}_{\mathbf{k}\uparrow}\\\hat{c}_{\mathbf{-k}\downarrow}\end{pmatrix} = \textrm{up},
\begin{pmatrix}\hat{c}^{\dagger}_{\mathbf{-k}\downarrow}\\\hat{c}_{\mathbf{k}\uparrow}\end{pmatrix} = \textrm{down},\\
\begin{pmatrix}\hat{c}_{\mathbf{k}\downarrow}\\\hat{c}^{\dagger}_{\mathbf{-k}\uparrow}\end{pmatrix} = \textrm{top},
\begin{pmatrix}\hat{c}_{\mathbf{-k}\uparrow}\\\hat{c}^{\dagger}_{\mathbf{k}\downarrow}\end{pmatrix} = \textrm{bottom}
\end{multline}
and it is the dynamics of these which gives rise to the interesting phenomena exhibited. The Yang-Mills interaction vertices are given by:\cite{Booth2018_1}
\begin{multline}
g_{(+,-,3)}\gamma^{\mu}\hat{W}_{\mu}(x)\psi=\begin{pmatrix}\gamma^{\mu}W^{3}_{\mu}(x)& \gamma^{\mu}W^{-}_{\mu}(x)\\ & \\ \gamma^{\mu}W^{+}_{\mu}(x)&-\gamma^{\mu}W^{3}_{\mu}(x)\end{pmatrix}\begin{pmatrix} \hat{c}^{\dagger}_{\mathbf{k}\uparrow}\\\hat{c}_{\mathbf{-k}\downarrow}\\\hat{c}^{\dagger}_{\mathbf{-k}\downarrow}\\\hat{c}_{\mathbf{k}\uparrow}\end{pmatrix}\\ g_{(+,-,3)}\gamma^{\mu}\hat{W}_{\mu}(x)\psi=\begin{pmatrix}\gamma^{\mu}W^{3}_{\mu}(x)&\gamma^{\mu}W^{-}_{\mu}(x)\\ & \\  \gamma^{\mu}W^{+}_{\mu}(x)&-\gamma^{\mu}W^{3}_{\mu}(x)\end{pmatrix}\begin{pmatrix} \hat{c}_{\mathbf{k}\downarrow}\\\hat{c}^{\dagger}_{\mathbf{-k}\uparrow}\\\hat{c}_{\mathbf{-k}\uparrow}\\\hat{c}^{\dagger}_{\mathbf{k}\downarrow}\end{pmatrix}
\end{multline}
The incoming spinor states can be grouped using spin as a ``gauge charge", i.e. a colour index, which gives the following structure:\cite{Booth2018_1}
\begin{figure}[h!]
\subfigure{\includegraphics[width=0.5\columnwidth]{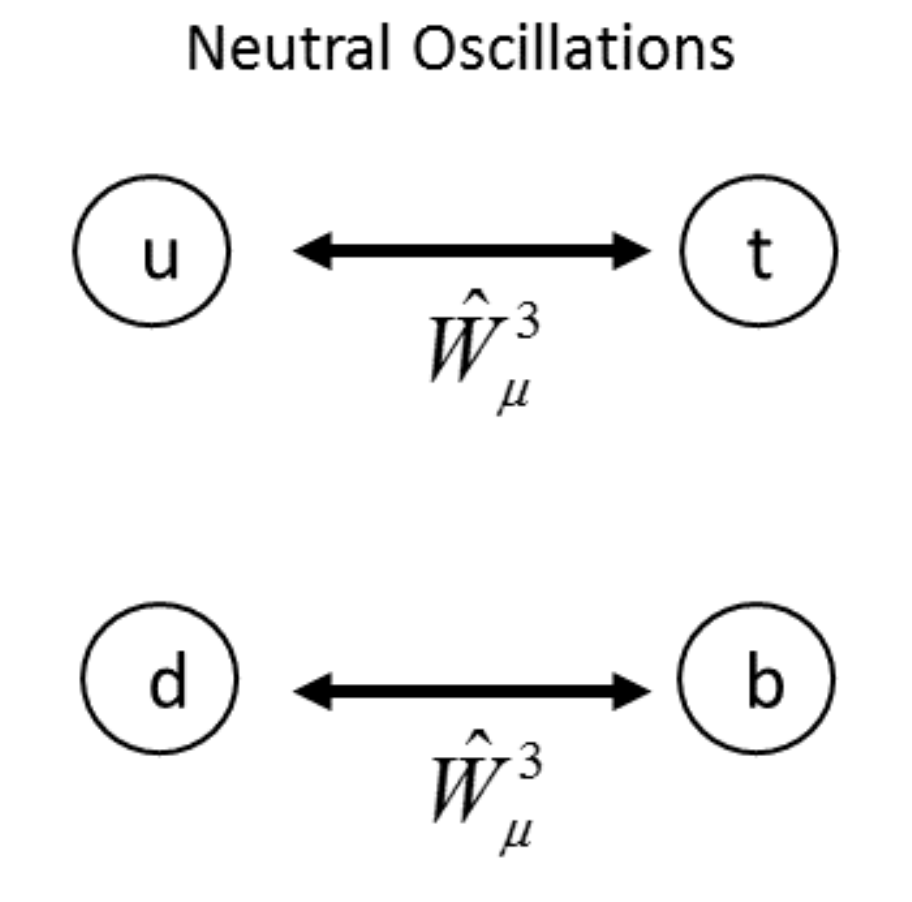}}{a)}\\
\subfigure{\includegraphics[width=0.55\columnwidth]{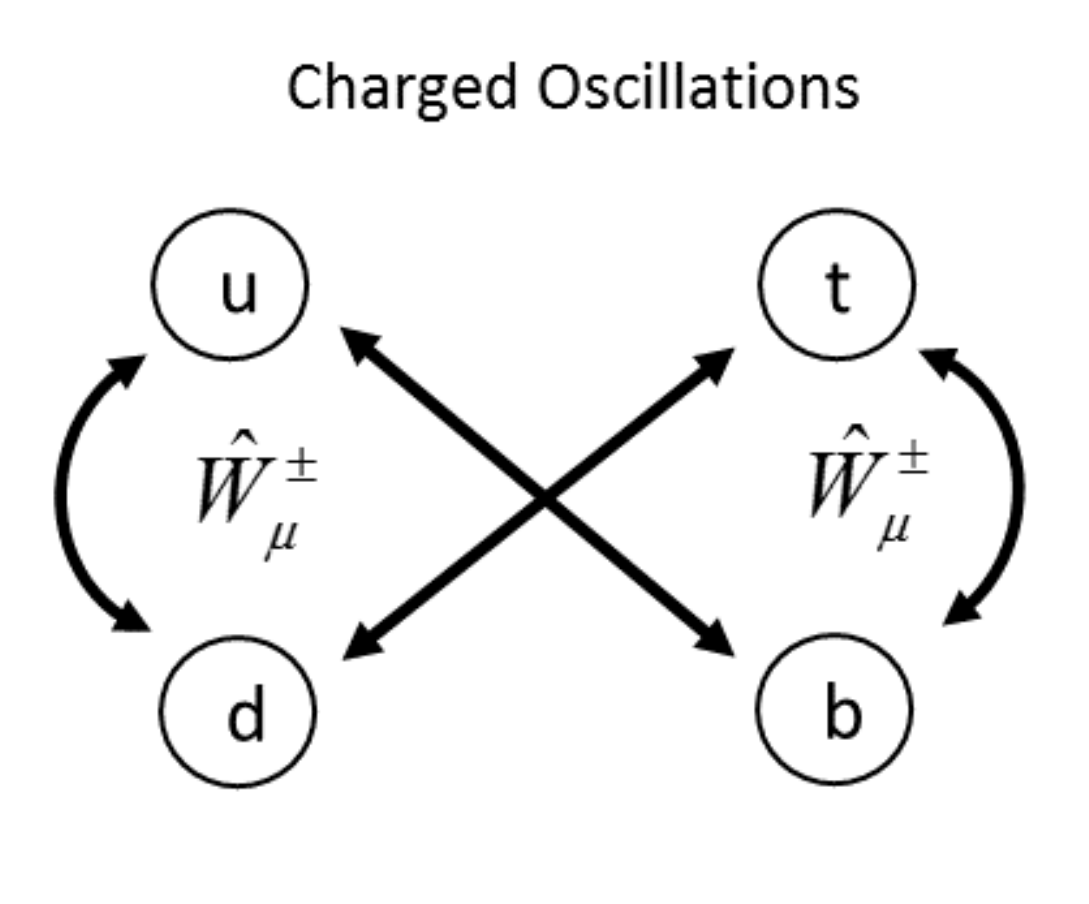}}{b)}
\caption{\raggedright{Schematic representation of the transformations enacted by the a) Neutral boson $W^{3}_{\mu}$ and b) the Charged bosons $W^{\pm}_{\mu}$}.}
\label{Schematics}
\end{figure}
\begin{center}
\begin{tabular}{c|cc}
\toprule
Colour &  \multicolumn{2}{c}{Flavour}\\
  & 1   & 2 \\
  \hline
\textbf{a} & up   & top \\
\textbf{b} & down   & bottom \\
\bottomrule
\end{tabular}
\end{center}
The actions of the bosons on the spinors is presented schematically in Figure (\ref{Schematics}), with the $\hat{W}^{3}_{\mu}$ field the neutral boson which doesn't couple to spin, while the $\hat{W}^{\pm}_{\mu}$ raising and lowering the spins. The most significant difference in this formalism is that making the ansatz that strong correlations result in the requirement of grouping the electron and hole states into 4-component Weyl spinors, and including phonon anharmonicity\cite{Booth2018_2} means that the currents which generate bosons are now comprised of double-stacked 4-component spinors with the SU(2) generators providing the outgoing boson structure.\cite{Booth2018_1} Thus the current becomes:
\begin{equation}
\hat{J}^{\mu} = \bar{\psi}\gamma^{\mu}\psi \rightarrow \bar{\psi}_{i}\hat{T}^{a}_{ij}\gamma^{\mu}\psi_{j}
\end{equation}
where $\hat{T}^{a}_{ij}$ is a generator of the group SU(2), i.e. a Pauli matrix. Thus the interaction between spinors in the context of the Yang-Mills interaction vertex to leading order goes as:\cite{Schwartz_Fierz}
\begin{equation}
-i\mathcal{M} = T^{a}_{ji}T^{a}_{kl}(ig_{a})^{2}\bar{\psi}_{j}\gamma^{\mu}\psi_{i}\frac{-i(g_{\mu\nu}-\frac{k_{\nu}k_{\mu}}{k^{2}})}{k^{2}}\bar{\psi}_{k}\gamma^{\nu}\psi_{l}
\label{Interaction}
\end{equation}
where for the transverse modes the term proportional to the momentum ($\frac{k_{\nu}k_{\mu}}{k^{2}}$) is zero, while for the longitudinal mode this term is non-zero, and positive and the $g_{\mu\nu}$ term is zero (this may seem a bit scrappy but the notation is more convenient). In this formalism, colour-anticolour pairs are Cooper pairs, so the input spinors for the interaction of equation (\ref{Interaction}) look like: 
\begin{multline}
u= \begin{pmatrix}\hat{c}^{\dagger}_{\mathbf{k\uparrow}}\\\hat{c}_{\mathbf{-k\downarrow}}\end{pmatrix},\quad \bar{u}=\begin{pmatrix}\hat{c}^{\dagger}_{\mathbf{-k}\downarrow}\\\hat{c}_{\mathbf{k\uparrow}}\end{pmatrix}, \\t =\begin{pmatrix}\hat{c}_{\mathbf{-k}\uparrow}\\\hat{c}^{\dagger}_{\mathbf{k\downarrow}}\end{pmatrix},\quad \bar{t}=\begin{pmatrix}\hat{c}_{\mathbf{k\downarrow}}\\\hat{c}^{\dagger}_{\mathbf{-k\uparrow}}\end{pmatrix}
\end{multline}
Plugging a Cooper pair in, for example $i = 1, k = 1$, or $i = 2, k = 2$ the colour prefactor can be calculated by the Fierz completeness relation:\cite{Schwartz_Fierz}
\begin{equation}
\sum_{a=1}^{3}T^{a}_{ji}T^{a}_{kl}=\frac{1}{2}(\delta_{il}\delta_{jk}-\frac{1}{2}\delta_{ij}\delta_{kl})=\begin{pmatrix}
\frac{1}{4}&0\\0&\frac{1}{2}
\end{pmatrix}_{jl}
\end{equation}
and is positive. For states with different colours, for example $i = 2, k = 1$, the colour factor is
\begin{equation}
\sum_{a=1}^{3}T^{a}_{ji}T^{a}_{kl}=\begin{pmatrix}
0&-\frac{1}{4}\\0&0
\end{pmatrix}_{jl}
\end{equation}
Therefore, different combinations of colours give different signs for the scattering effective potential. Since the scattering factor for colour-anticolour pairs is positive, the potential between them is \textit{attractive}, while for different colours the sign is negative, and therefore the potential is repulsive.

Therefore since the scattering of colour singlets by phonons in a Yang-Mills theory gives an attractive interaction, the Bethe-Salpeter equation will give an instability for the formation of bound states (Cooper Pairs) similar to the Fr{\"o}hlich interaction in conventional superconductors,\cite{Altland2006_2} but without the requirement of the propagators being off-shell. Thus the fairly tortuous derivation of Bardeen and Pines\cite{Bardeen1955} is unnecessary in Yang-Mills theory, colour singlet states are automatically confined, and this interaction does not require the existence of quasiparticles, but \textit{does} require the existence of a many-body ground state of electron and hole pairs to act as a source for the Yang-Mills bosons.\cite{Booth2018_1}

The author acknowledges the support of the ARC Centre of Excellence in Exciton Science (CE170100026). Correspondence and requests for materials should be addressed to JMB, email: jamie.booth@rmit.edu.au
\bibliography{C:/Local_Disk/GWApproximation/Bibliography/library}

\begin{thebibliography}{15}%
\makeatletter
\providecommand \@ifxundefined [1]{%
 \@ifx{#1\undefined}
}%
\providecommand \@ifnum [1]{%
 \ifnum #1\expandafter \@firstoftwo
 \else \expandafter \@secondoftwo
 \fi
}%
\providecommand \@ifx [1]{%
 \ifx #1\expandafter \@firstoftwo
 \else \expandafter \@secondoftwo
 \fi
}%
\providecommand \natexlab [1]{#1}%
\providecommand \enquote  [1]{``#1''}%
\providecommand \bibnamefont  [1]{#1}%
\providecommand \bibfnamefont [1]{#1}%
\providecommand \citenamefont [1]{#1}%
\providecommand \href@noop [0]{\@secondoftwo}%
\providecommand \href [0]{\begingroup \@sanitize@url \@href}%
\providecommand \@href[1]{\@@startlink{#1}\@@href}%
\providecommand \@@href[1]{\endgroup#1\@@endlink}%
\providecommand \@sanitize@url [0]{\catcode `\\12\catcode `\$12\catcode
  `\&12\catcode `\#12\catcode `\^12\catcode `\_12\catcode `\%12\relax}%
\providecommand \@@startlink[1]{}%
\providecommand \@@endlink[0]{}%
\providecommand \url  [0]{\begingroup\@sanitize@url \@url }%
\providecommand \@url [1]{\endgroup\@href {#1}{\urlprefix }}%
\providecommand \urlprefix  [0]{URL }%
\providecommand \Eprint [0]{\href }%
\providecommand \doibase [0]{http://dx.doi.org/}%
\providecommand \selectlanguage [0]{\@gobble}%
\providecommand \bibinfo  [0]{\@secondoftwo}%
\providecommand \bibfield  [0]{\@secondoftwo}%
\providecommand \translation [1]{[#1]}%
\providecommand \BibitemOpen [0]{}%
\providecommand \bibitemStop [0]{}%
\providecommand \bibitemNoStop [0]{.\EOS\space}%
\providecommand \EOS [0]{\spacefactor3000\relax}%
\providecommand \BibitemShut  [1]{\csname bibitem#1\endcsname}%
\let\auto@bib@innerbib\@empty
\bibitem [{\citenamefont {Bednorz}\ and\ \citenamefont
  {Mueller}(1986)}]{Bednorz1986}%
  \BibitemOpen
  \bibfield  {author} {\bibinfo {author} {\bibfnamefont {J.~G.}\ \bibnamefont
  {Bednorz}}\ and\ \bibinfo {author} {\bibfnamefont {K.~A.}\ \bibnamefont
  {Mueller}},\ }\href@noop {} {\bibfield  {journal} {\bibinfo  {journal} {Z.
  Phys. B}\ }\textbf {\bibinfo {volume} {64}},\ \bibinfo {pages} {189}
  (\bibinfo {year} {1986})}\BibitemShut {NoStop}%
\bibitem [{\citenamefont {Kamihara}\ \emph {et~al.}(2006)\citenamefont
  {Kamihara}, \citenamefont {Hiramatsu}, \citenamefont {Hirano}, \citenamefont
  {Kawamura}, \citenamefont {Yanagi}, \citenamefont {Kamiya},\ and\
  \citenamefont {Hosono}}]{Kamihara2006}%
  \BibitemOpen
  \bibfield  {author} {\bibinfo {author} {\bibfnamefont {Y.}~\bibnamefont
  {Kamihara}}, \bibinfo {author} {\bibfnamefont {H.}~\bibnamefont {Hiramatsu}},
  \bibinfo {author} {\bibfnamefont {M.}~\bibnamefont {Hirano}}, \bibinfo
  {author} {\bibfnamefont {R.}~\bibnamefont {Kawamura}}, \bibinfo {author}
  {\bibfnamefont {H.}~\bibnamefont {Yanagi}}, \bibinfo {author} {\bibfnamefont
  {T.}~\bibnamefont {Kamiya}}, \ and\ \bibinfo {author} {\bibfnamefont
  {H.}~\bibnamefont {Hosono}},\ }\href {\doibase 10.1021/ja063355c} {\bibfield
  {journal} {\bibinfo  {journal} {Journal of the American Chemical Society}\
  }\textbf {\bibinfo {volume} {128}},\ \bibinfo {pages} {10012} (\bibinfo
  {year} {2006})},\ \Eprint {http://arxiv.org/abs/arXiv:0803.3286v1}
  {arXiv:arXiv:0803.3286v1} \BibitemShut {NoStop}%
\bibitem [{\citenamefont {Mann}(2011)}]{Mann2011}%
  \BibitemOpen
  \bibfield  {author} {\bibinfo {author} {\bibfnamefont {A.}~\bibnamefont
  {Mann}},\ }\href {\doibase 10.1038/475280a} {\bibfield  {journal} {\bibinfo
  {journal} {Nature}\ }\textbf {\bibinfo {volume} {475}},\ \bibinfo {pages}
  {280} (\bibinfo {year} {2011})}\BibitemShut {NoStop}%
\bibitem [{\citenamefont {Sachdev}\ and\ \citenamefont
  {Chowdhury}(2016)}]{Sachdev2016}%
  \BibitemOpen
  \bibfield  {author} {\bibinfo {author} {\bibfnamefont {S.}~\bibnamefont
  {Sachdev}}\ and\ \bibinfo {author} {\bibfnamefont {D.}~\bibnamefont
  {Chowdhury}},\ }\href {\doibase 10.1093/ptep/ptw110} {\bibfield  {journal}
  {\bibinfo  {journal} {Progress of Theoretical and Experimental Physics}\
  }\textbf {\bibinfo {volume} {2016}},\ \bibinfo {pages} {1} (\bibinfo {year}
  {2016})},\ \Eprint {http://arxiv.org/abs/1605.03579} {arXiv:1605.03579}
  \BibitemShut {NoStop}%
\bibitem [{\citenamefont {Hashimoto}\ \emph {et~al.}(2014)\citenamefont
  {Hashimoto}, \citenamefont {Vishik}, \citenamefont {He}, \citenamefont
  {Devereaux},\ and\ \citenamefont {Shen}}]{Hashimoto2014}%
  \BibitemOpen
  \bibfield  {author} {\bibinfo {author} {\bibfnamefont {M.}~\bibnamefont
  {Hashimoto}}, \bibinfo {author} {\bibfnamefont {I.~M.}\ \bibnamefont
  {Vishik}}, \bibinfo {author} {\bibfnamefont {R.-H.}\ \bibnamefont {He}},
  \bibinfo {author} {\bibfnamefont {T.~P.}\ \bibnamefont {Devereaux}}, \ and\
  \bibinfo {author} {\bibfnamefont {Z.-X.}\ \bibnamefont {Shen}},\ }\href
  {\doibase 10.1038/nphys3009} {\bibfield  {journal} {\bibinfo  {journal} {Nat.
  Phys.}\ }\textbf {\bibinfo {volume} {10}},\ \bibinfo {pages} {483} (\bibinfo
  {year} {2014})}\BibitemShut {NoStop}%
\bibitem [{\citenamefont {Anderson}(1987)}]{Anderson1987}%
  \BibitemOpen
  \bibfield  {author} {\bibinfo {author} {\bibfnamefont {P.~W.}\ \bibnamefont
  {Anderson}},\ }\href {\doibase 10.1126/science.235.4793.1196} {\bibfield
  {journal} {\bibinfo  {journal} {Science}\ }\textbf {\bibinfo {volume}
  {235}},\ \bibinfo {pages} {1} (\bibinfo {year} {1987})}\BibitemShut {NoStop}%
\bibitem [{\citenamefont {Miyake}\ \emph {et~al.}(1986)\citenamefont {Miyake},
  \citenamefont {Schmitt-Rink},\ and\ \citenamefont {Varma}}]{Miyake1986}%
  \BibitemOpen
  \bibfield  {author} {\bibinfo {author} {\bibfnamefont {K.}~\bibnamefont
  {Miyake}}, \bibinfo {author} {\bibfnamefont {S.}~\bibnamefont
  {Schmitt-Rink}}, \ and\ \bibinfo {author} {\bibfnamefont {C.~M.}\
  \bibnamefont {Varma}},\ }\href {\doibase 10.1103/PhysRevB.34.6554} {\bibfield
   {journal} {\bibinfo  {journal} {Physical Review B}\ }\textbf {\bibinfo
  {volume} {34}},\ \bibinfo {pages} {6554} (\bibinfo {year}
  {1986})}\BibitemShut {NoStop}%
\bibitem [{\citenamefont {Scalapino}\ \emph {et~al.}(1986)\citenamefont
  {Scalapino}, \citenamefont {Loh},\ and\ \citenamefont
  {Hirsch}}]{Scalapino1986}%
  \BibitemOpen
  \bibfield  {author} {\bibinfo {author} {\bibfnamefont {D.~J.}\ \bibnamefont
  {Scalapino}}, \bibinfo {author} {\bibfnamefont {E.}~\bibnamefont {Loh}}, \
  and\ \bibinfo {author} {\bibfnamefont {J.~E.}\ \bibnamefont {Hirsch}},\
  }\href {\doibase 10.1103/PhysRevB.34.8190} {\bibfield  {journal} {\bibinfo
  {journal} {Phys. Rev. B}\ }\textbf {\bibinfo {volume} {34}},\ \bibinfo
  {pages} {8190} (\bibinfo {year} {1986})}\BibitemShut {NoStop}%
\bibitem [{\citenamefont {Lee}\ \emph {et~al.}(2006)\citenamefont {Lee},
  \citenamefont {Nagaosa},\ and\ \citenamefont {Wen}}]{Lee2006}%
  \BibitemOpen
  \bibfield  {author} {\bibinfo {author} {\bibfnamefont {P.~A.}\ \bibnamefont
  {Lee}}, \bibinfo {author} {\bibfnamefont {N.}~\bibnamefont {Nagaosa}}, \ and\
  \bibinfo {author} {\bibfnamefont {X.-G.}\ \bibnamefont {Wen}},\ }\href
  {\doibase 10.1103/RevModPhys.78.17} {\bibfield  {journal} {\bibinfo
  {journal} {Rev. Mod. Phys.}\ }\textbf {\bibinfo {volume} {78}},\ \bibinfo
  {pages} {17} (\bibinfo {year} {2006})}\BibitemShut {NoStop}%
\bibitem [{\citenamefont {Keimer}\ \emph {et~al.}(2015)\citenamefont {Keimer},
  \citenamefont {Kivelson}, \citenamefont {Norman}, \citenamefont {Uchida},\
  and\ \citenamefont {Zaanen}}]{Keimer2015}%
  \BibitemOpen
  \bibfield  {author} {\bibinfo {author} {\bibfnamefont {B.}~\bibnamefont
  {Keimer}}, \bibinfo {author} {\bibfnamefont {S.~A.}\ \bibnamefont
  {Kivelson}}, \bibinfo {author} {\bibfnamefont {M.~R.}\ \bibnamefont
  {Norman}}, \bibinfo {author} {\bibfnamefont {S.}~\bibnamefont {Uchida}}, \
  and\ \bibinfo {author} {\bibfnamefont {J.}~\bibnamefont {Zaanen}},\ }\href
  {\doibase 10.1038/nature14165} {\bibfield  {journal} {\bibinfo  {journal}
  {Nature}\ }\textbf {\bibinfo {volume} {518}},\ \bibinfo {pages} {179}
  (\bibinfo {year} {2015})}\BibitemShut {NoStop}%
\bibitem [{\citenamefont {Booth}\ and\ \citenamefont
  {Russo}(2018)}]{Booth2018_1}%
  \BibitemOpen
  \bibfield  {author} {\bibinfo {author} {\bibfnamefont {J.~M.}\ \bibnamefont
  {Booth}}\ and\ \bibinfo {author} {\bibfnamefont {S.~P.}\ \bibnamefont
  {Russo}},\ }\href@noop {} {\  (\bibinfo {year} {2018})},\ \Eprint
  {http://arxiv.org/abs/1808.05769v2} {arXiv:1808.05769v2} \BibitemShut
  {NoStop}%
\bibitem [{\citenamefont {Booth}(2018)}]{Booth2018_2}%
  \BibitemOpen
  \bibfield  {author} {\bibinfo {author} {\bibfnamefont {J.~M.}\ \bibnamefont
  {Booth}},\ }\href {http://arxiv.org/abs/1810.03273} {\  (\bibinfo {year}
  {2018})},\ \Eprint {http://arxiv.org/abs/1810.03273} {arXiv:1810.03273}
  \BibitemShut {NoStop}%
\bibitem [{\citenamefont {Schwartz}(2014)}]{Schwartz_Fierz}%
  \BibitemOpen
  \bibfield  {author} {\bibinfo {author} {\bibfnamefont {M.~D.}\ \bibnamefont
  {Schwartz}},\ }\href@noop {} {\emph {\bibinfo {title} {{Quantum Field Theory
  and the Standard Model}}}}\ (\bibinfo  {publisher} {Cambridge University
  Press},\ \bibinfo {address} {Cambridge},\ \bibinfo {year} {2014})\ p.\
  \bibinfo {pages} {512}\BibitemShut {NoStop}%
\bibitem [{\citenamefont {Altland}\ and\ \citenamefont
  {Simons}(2006)}]{Altland2006_2}%
  \BibitemOpen
  \bibfield  {author} {\bibinfo {author} {\bibfnamefont {A.}~\bibnamefont
  {Altland}}\ and\ \bibinfo {author} {\bibfnamefont {B.}~\bibnamefont
  {Simons}},\ }in\ \href@noop {} {\emph {\bibinfo {booktitle} {Condensed Matter
  Field Theory}}}\ (\bibinfo  {publisher} {Cambridge University Press},\
  \bibinfo {address} {Cambridge},\ \bibinfo {year} {2006})\ Chap.~\bibinfo
  {chapter} {6}, p.\ \bibinfo {pages} {269}\BibitemShut {NoStop}%
\bibitem [{\citenamefont {Bardeen}\ and\ \citenamefont
  {Pines}(1955)}]{Bardeen1955}%
  \BibitemOpen
  \bibfield  {author} {\bibinfo {author} {\bibfnamefont {J.}~\bibnamefont
  {Bardeen}}\ and\ \bibinfo {author} {\bibfnamefont {D.}~\bibnamefont
  {Pines}},\ }\href {\doibase 10.1103/PhysRev.99.1140} {\bibfield  {journal}
  {\bibinfo  {journal} {Phys. Rev.}\ }\textbf {\bibinfo {volume} {99}},\
  \bibinfo {pages} {1140} (\bibinfo {year} {1955})}\BibitemShut {NoStop}%
\end{thebibliography}%
\end{document}